\documentclass[a4paper,11pt]{article}

\pdfoutput=1 

\usepackage{./style_files/jheppub} 

\usepackage[T1]{fontenc} 

\title{\boldmath Multi-critical Points in Black Hole Phase Transitions}

\author[a]{Masoumeh Tavakoli}
\author[a]{Jerry Wu}
\author[a,b,1]{Robert B. Mann,\note{Corresponding author.}}

\affiliation[a]{Department of Physics and Astronomy, University of Waterloo, \\ Waterloo, Ontario, Canada N2L 3G1}
\affiliation[b]{Perimeter Institute for Theoretical Physics, \\ 31 Caroline St N, Waterloo, Ontario, Canada N2L 2Y5}

\emailAdd{tavakoli.phy@gmail.com}
\emailAdd{yq4wu@uwaterloo.ca} 
\emailAdd{rbmann@uwaterloo.ca}

\abstract{We present  the first examples  in black hole thermodynamics of multicritical phase transitions, in which more than three distinct black hole phases merge at a critical point.  Working in the context of non-linear electrodynamics, we explicitly present  examples of black hole quadruple and quintuple points, and demonstrate how $n$-tuple critical points can be obtained.  Our results indicate that black holes can have multiple phases beyond the three types observed so far, resembling the behaviour of multicomponent chemical systems.  We discuss the interpretation of our results in the context of the Gibbs Phase Rule.
}

\usepackage{braket}
\usepackage{color}
\usepackage[dvipsnames]{xcolor}

\newcommand{\be}{\begin{equation}}
\newcommand{\ee}{\end{equation}}
\newcommand{\ba}{\begin{eqnarray}}
\newcommand{\ea}{\end{eqnarray}}

\usepackage{colonequals}

\usepackage[utf8]{inputenc}
\usepackage[T1]{fontenc}
\usepackage{fancyhdr}
\usepackage{lipsum}
\usepackage{graphicx}
\usepackage{titlesec}
\usepackage{appendix}
\usepackage[sectionbib]{chapterbib}
\usepackage[breakwords]{truncate}
\usepackage{lastpage}
\usepackage{amsmath}
\usepackage{hyperref}
\usepackage{amssymb}
\usepackage[font={small}]{caption}
\usepackage{physics}

\makeatletter
\gdef\@fpheader{}
\makeatother

\begin{document}
\maketitle
\flushbottom

\newpage

\section{Introduction}

The importance of black hole  thermodynamics   in providing 
clues about the nature of quantum  gravity cannot be underestimated.  Much has been learned from studying asymptotically anti de Sitter (AdS) black holes, beginning with the pioneering work of Hawking and Page who demonstrated the existence of a phase transition between thermal radiation and a large AdS black hole~\cite{Hawking:1982dh}.  Such black holes can be in thermal equilibrium with their Hawking radiation, and the 
Hawking--Page phase transition  corresponds to the confinement/deconfinement of the dual quark gluon plasma \cite{Witten:1998zw}.

More generally,  the cosmological constant can be regarded as a thermodynamic variable corresponding to pressure in the first law~\cite{Caldarelli:1999xj, Kastor:2009wy, Cvetic:2010jb} 
(arising, e.g. as a $(d-1)$-form gauge field \cite{Creighton:1995au}).   Within this context, the black hole mass takes on the interpretation of enthalpy and a rich variety of thermodynamic phase behaviour has been shown to emerge. The Hawking-Page transition can be understood as a solid-liquid transition \cite{Kubiznak:2014zwa}, and  black holes have been shown to exhibit  Van der Waals \cite{Kubiznak:2012wp}, re-entrant \cite{Altamirano:2013ane}, superfluid \cite{Hennigar:2016xwd}, and polymer-type phase transitions \cite{Dolan:2014vba}, snapping transitions  \cite{Abbasvandi:2018vsh} for accelerating black holes \cite{Anabalon:2018qfv}, and universal scaling behaviour of the Ruppeiner curvature \cite{Wei:2019yvs}.  This panoply of behaviour has suggested a 
molecular interpretation of the underlying constituent degrees of freedom \cite{Wei:2019uqg}. 
  For these reasons this subdiscipline has come to be called {\em Black Hole Chemistry} \cite{Kubiznak:2016qmn}.  

Triple points have been observed for quite some time \cite{Altamirano:2013uqa, Wei:2014hba,Frassino:2014pha}, and a recent interpretation of the microstructure of black holes at  such points was recently proposed \cite{Wei:2021krr}. However multicritical points -- in which many phases merge together at a single critical point -- have never been observed.   

We present here the first examples of black hole multicritical points in the context of non-linear electrodynamics in 4-dimensional Einstein gravity.  We find that regions exist within the parameter space of such theories in which many distinct black hole phases exist for a range of sufficiently large pressure.  Transitions between each such phase are of 1st-order, and the coexistence lines for each terminate in distinct 2nd-order critical points.  As the pressure is lowered to a certain critical value, all these phases merge into a single multicritical point.  Below this critical pressure only two distinct phases exist, separated by a 1st-order phase transition between the largest and the smallest possible black holes allowed.   

We explicitly illustrate this for a black hole quadruple point.  At high pressures only a single black hole phase exists.  As the pressure is lowered, first two, then three, and finally four distinct black hole phases emerge, each distinguished by their size.  As the pressure is further lowered, all four phases merge at a single tetra-critical point (a quadruple point).  For pressures below this point there are only two distinct phases -- the largest and smallest possible black holes -- separated by a first-order phase transition.

Our results provide the first indication that black holes can behave like multicomponent systems in nature that exhibit similar multicritical behaviour \cite{PhysRevLett.125.127803}.  
We find that an $n$-tuple critical point for a charged black hole requires $2n-1$ conjugate pairs in Power-Maxwell theory.  The Gibbs Phase Rule then implies that there are $n$ degrees of freedom at this multicritical point, whose implications for the microstructure of black holes has yet to be clarified.  We preserve up to 10 significant digits for certain thermodynamic parameters as minuscule changes can drastically alter the phase behaviour in some cases.

\section{Black Hole Thermodynamics in
Einstein-Power-Maxwell Theory}

In the {\em extended thermodynamic phase space} of black hole chemistry \cite{Kastor:2009wy,Kubiznak:2016qmn}, the relation
\be\label{P}
P=-\frac{\Lambda}{8\pi G}\,,\quad \Lambda=-\frac{(D-1)(D-2)}{2 l^2}\,, 
\ee
is posited between  a  (negative) cosmological constant $\Lambda$ and the thermodynamic pressure $P$, with 
 $l$ is the radius of the $D$-dimensional AdS space and $G$ the (dimensionful) Newton  gravitational constant;
 we set $\hbar=c=1$. The black hole mass $M$ is  interpreted as thermodynamic enthalpy rather than internal energy. The   first law of thermodynamics and   corresponding Smarr relation  are
\ba
\delta M&=&T\delta S+V\delta P+\phi \delta Q+\Omega \delta J\,,\label{flaw}\\
M&=&\frac{D-2}{D-3}(TS+\Omega J)+\phi Q-\frac{2}{D-3}PV\,,\label{smarr}
\ea 
for a black hole of charge $Q$, surface gravity $\kappa$, angular momentum $J$, and area $A$ in $D$-dimensional Einstein gravity, 
where entropy $S$ and temperature $T$  are
\be
S=\frac{A}{4G}\,,\quad T=\frac{\kappa}{2\pi}\,, 
\ee
with $\phi,\Omega$ the respective conjugates to $Q,J$, and  where
\be\label{V}
V=\Bigl(\frac{\partial M}{\partial P}\Bigr)_{S,Q,J}\,
\ee
is the thermodynamic volume  conjugate to $P$ \cite{Kastor:2009wy, Cvetic:2010jb,Dolan:2010ha}.

We consider  a general form of non-linear electrodynamics minimally coupled to $D=4$ Einstein gravity \cite{Gao:2021kvr}. The action of this theory is  
\begin{equation}\label{action}
	S=\int d^4x \sqrt{-g}\left(R-2 \Lambda -\sum_{i=1}^{N} \alpha _i (F^2)^i \right),
\end{equation}
with $F^2 \equiv F_{\mu \nu }F^{\mu \nu}$ and $F_{\mu \nu}\equiv \nabla_{\mu} A_\nu - \nabla _\nu A_\nu$, where
  $R$ is the Ricci scalar. The $ \alpha _i$ are dimensional coupling constants ( $[\alpha_i]=L^{2(i-1)}$)  and $A_\mu$ is the U(1) Maxwell field. We recover Einstein-Maxwell theory when $\alpha_1=1$ and $\alpha_i (i>1) = 0$. 
  
 Writing $L_{EM} = -\sum_{i=1}^{N} \alpha _i (F^2)^i $ we obtain 
the Einstein Power-Maxwell equations  
\begin{align}
	&G_{\mu \nu}=-2 \frac{dL_{EM}}{dF^2} F_\mu^{\; \lambda}F _{\nu \lambda}+\frac{1}{2} g_{\mu \nu } L_{EM} \\
&	\nabla_{\mu}\left( \frac{dL_{EM}}{dF^2} F^{\mu \nu }\right)=0
\end{align}
by varying the action with respect to the metric and gauge field. 

The following ansatz 
\begin{align}
	ds^2 &= - U(r) dt^2+\frac{1}{U(r)}dr^2+ r^2 d \Omega^2_2 \nonumber\\
	A_\mu &= [\Phi (r), 0, 0, 0]
\label{ansatz}
\end{align}
yields the field equations
\begin{align}
&\left(r (U(r) -1)\right)^\prime + r^2\Lambda \nonumber \\ & \qquad \qquad- r^2 \sum_{n=1}^N \left(n-\frac{1}{2}\right)\alpha_n \left( -2 (\Phi^\prime)^2\right)^n = 0 
\label{Ein5} \\
& \qquad \quad \frac{1}{2} r^2 \sum_{n=1}^N n \alpha_n \left( -2 (\Phi^\prime)^2\right)^n - Q(\Phi^\prime) = 0
\label{Max5}
\end{align}
with  $Q$  an integration constant corresponding to the electric charge of the  black hole, and the prime denoting a derivative with respect to $r$.
If $\Lambda=0$, asymptotically flat black holes with multiple horizons can be obtained
 \cite{Gao:2021kvr}   provided the coupling constants $\alpha _i $ are appropriately chosen.  Writing $\Lambda=-3/l^2$, this  result is straightforwardly generalized to the asymptotically AdS case by writing
\begin{align}
	&\Phi = \sum_{i=1}^{K} b_i r^{-i}  \quad U =1+\sum_{i=1}^{K} c_i r^{-i}+\frac{r^2}{l^2} \label{eq12_2}
\end{align}
where the field equations imply
\begin{equation}
   c_1= - 2M
    \qquad
  c_i=\frac{4Q}{i+2}b_{i-1},
    \qquad \text{for}  \qquad
    i>1
  \end{equation}
setting $\alpha_1 =1$ without loss of generality, with
\begin{align}
	&b_1=Q \qquad 
	b_5=\frac{4}{5}Q^3\alpha_2
	\qquad
	b_9=\frac{4}{3}Q^5(4\alpha_2^2-\alpha_3) 
	\\
	&b_{13} = \frac{32}{13}Q^7(24 \alpha_2^3-12 \alpha_3\alpha_2+\alpha_4)
	\\
	&b_{17}=\frac{80}{17}Q^9(1
	76\alpha_2^4-132 \alpha_2^2\alpha_3+16\alpha_4\alpha_2+9\alpha_3^2-\alpha_5)
	\\
	&b_{21}=\frac{64}{7}Q^{11}(1456\alpha_2^5+234\alpha_3^2\alpha_2+208\alpha_4\alpha_2^2-24\alpha_4\alpha_3
	\nonumber\\
	&\qquad -1456\alpha_2^3\alpha_3-20\alpha_5\alpha_2+\alpha_6
	)
	\\
	&...=......, \nonumber
\end{align}
If  all $\alpha_i$ for $(i>1)$ are set to zero, we recover 
\begin{align}
	\Phi =\frac{Q}{r}  \qquad U =\frac{r^2}{l^2}+1-\frac{2M}{r}+\frac{Q^2}{r^2} 
\end{align}
which is the usual Reissner-Nordstrom AdS-black hole having two horizons. 
For appropriate choices of 
non-vanishing
$\alpha_i$ with $i \leq n-1$, $n$ distinct horizons can be obtained.
 
The parameter $M$ corresponds to the mass of the black hole, namely the
conserved charge associated with the timelike Killing vector $\xi = \partial_t$ of the metric \eqref{ansatz}.
To demonstrate this,
we employ the Ashtekar--Das definition of conformal
mass \cite{Ashtekar:1999jx,Das:2000cu}, which extracts the mass
of the spacetime near the boundary via conformal regularization.
To do so, we carry out a conformal transformation on the metric  \eqref{ansatz} 
$\bar{g}_{\mu\nu} = {\bar{\Omega}}^{2}g_{\mu\nu}$,
to remove the divergence near the boundary $r\to\infty$.  We  then  obtain a
conserved charge by  integrating the conserved current
 \be
Q(\xi )=\frac{\ell}{8\pi}\lim_{\bar{\Omega} \rightarrow 0}\oint
\frac{\ell^{2}}{\bar{\Omega}}N^{\alpha }N^{\beta }
\bar{C}^{\nu}{}_{\alpha \mu \beta }
\xi _{\nu }d\bar{S}^{\mu } \nonumber
\ee
composed of the Weyl tensor of the conformal metric,
$\bar{C}^{\mu }{}_{\alpha \nu \beta }$, where
$$
N_{\mu }=\partial _{\mu }{\bar{\Omega}} 
$$
is the  normal
to the boundary,  and
\begin{equation} 
d\bar{S}_{\mu }=\delta^{\tau }_{\mu } \ell^{2} (d\cos\theta) d\phi  \nonumber
\end{equation}
is spacelike surface element
tangent to $ \bar{\Omega}=0$.

Although the conformal completion is not unique, the
charge $Q(\xi )$   is independent of the choice of conformal completion.
For simplicity, we can choose
   $\bar{\Omega}=\ell\Omega r^{-1} $.  This yields
$$
Q(\partial _{\tau})= \lim_{r\to\infty}\left( M - \frac{Q^2}{r} - \frac{14 b_5}{3 r^5} -11 \frac{b_9}{r^9} + \cdots + {\cal O}\left(\frac{b_n}{r^n} \right) \right)
$$   
and so  we obtain
\begin{equation} 
 Q(\partial _{\tau})= M \nonumber
\end{equation}
for the mass.  This will hold for any asymptotically AdS solution to the field equations \eqref{Ein5}, \eqref{Max5} since every $b_i$ coefficient in
the metric function $U$ falls off faster than $1/r$.

\begin{figure*}[t!]
    \centering
	\includegraphics[width=0.46\textwidth]{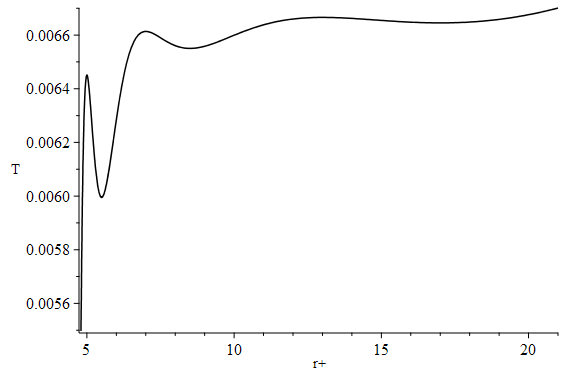}
	\includegraphics[width=0.5\textwidth]{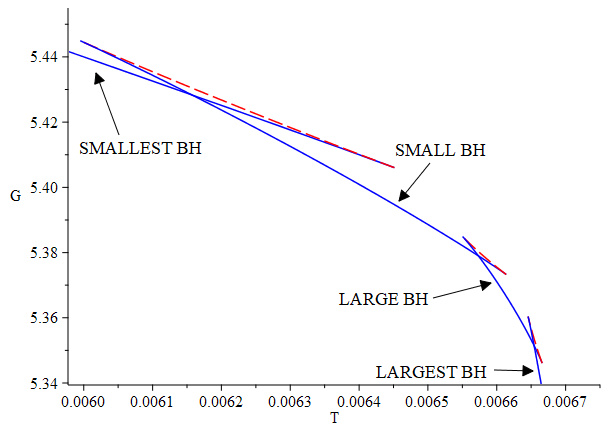}

	\caption{\textbf{Three stable first order phase transitions.} $P = 7.82\times 10^{-5}, Q = 6.623, \alpha_2 = -21.63694203, \alpha_3 = 1493.535254, \alpha_4 = -148046.3896, \alpha_5 = 1.759261993 \times 10^7, \alpha_6 = -2.332423991\times 10^9, \alpha_7 = 3.327781293 \times 10^{11}.$ Coupling constants are rounded to 10 digits. \textit{Left.} Inflections in $T(r_+)$ occur at increasing temperatures. \textit{Right.} Three separated swallowtails in the Gibbs free energy, indicating three first order phase transitions. 
	}
	\label{fig:3separated} 
\end{figure*}

Any given solution depends on a finite number of parameters that can be indexed either by $K$ (the number of parameters in the 
solution) or $N$  (the number of $\alpha_i$ couplings in the action), along with $M$ and $Q$.  In the former case the expressions for
the metric and gauge field are explicitly given by \eqref{eq12_2} whereas in the latter case these expressions are implicitly determined.
In either case  appropriate choices of a finite number of $\alpha_i$ exist that yield multicritical thermodynamic behaviour, with the value of $K$ or $N$
determining the maximal degree of multicriticality.   

\section{Multicritical Behaviour}

We begin by illustrating the existence of a quadruple point, which can be obtained 
for $\alpha_i \neq 0$ with
$i\leq 7$, sufficient to support up to eight black hole horizons \cite{Gao:2021kvr}. Explicitly
\begin{align}
	\Phi = & \frac{Q}{r} + \frac{b_5}{r^5} + \frac{b_9}{r^9} + \frac{b_{13}}{r^{13}} + \frac{b_{17}}{r^{17}} + \frac{b_{21}}{r^{21}} + \frac{b_{25}}{r^{25}} \\
	U =& 1 - \frac{2M}{r} + \frac{Q^2}{r^2} + \frac{b_5 Q}{2r^6}  + \frac{b_9 Q}{3 r^{10}}  + \frac{b_{13} Q}{4 r^{14}}   \notag \\
	& + \frac{b_{17} Q}{5 r^{18}}  + \frac{b_{21} Q}{6 r^{22}}   + \frac{b_{25} Q}{7 r^{26}}   + \frac{r^2}{l^2} 
\end{align}
with $b_i = 0$ for $i>25$. 

The thermodynamic quantities are given by
\begin{align}
	T & =\frac{1}{4\pi r_{+}}
	\biggl( 1 + \frac{3 r_{+}^{2}}{l^2} - \frac{Q^2}{r_{+}^2} - \frac{5 b_5 Q}{2 r_{+}^{6}} - \frac{3 b_9 Q}{r_{+}^{10}} \notag \\ 
	&  - \frac{13 b_{13} Q}{4 r_{+}^{14}} - \frac{17 b_{17} Q}{5 r_{+}^{18}} - \frac{7 b_{21} Q}{2 r_{+}^{22}} - \frac{25 b_{25} Q}{7 r_{+}^{26}} \biggr) \\
	S &= \pi r_+^2 
	\qquad
	V =\frac{4}{3}\pi r^3_+
	\qquad
	P =\frac{3}{8 \pi l ^2}
\end{align}
in Planckian units 
$l^2_P =\frac{G\hbar}{c^3} $ \cite{Kubiznak:2016qmn},
with  equation of state 
\begin{align}
    P &= \frac{T}{2r_{+}} - \frac{1}{8 \pi r_{+}^2} + \frac{Q^2}{8 \pi r_{+}^4}  + \frac{5 b_5 Q}{16 \pi r_{+}^8} + \frac{3 b_9 Q}{8 \pi r_{+}^{12}} \notag \\ & 
    \qquad + \frac{13 b_{13} Q}{32 \pi r_{+}^{16}} + \frac{17 b_{17} Q}{40 \pi r_{+}^{20}} + \frac{7 b_{21} Q}{16 \pi r_{+}^{24}} + \frac{25 b_{25} Q}{56 \pi r_{+}^{28}} 
\end{align}
and Gibbs free energy $G=M-TS$.

By carefully choosing the locations of the local extrema in $T(r_+)$, it is possible to produce
three separate swallowtails in the Gibbs free energy, characteristic of three stable first order phase transitions   (Fig.~\ref{fig:3separated}). 
This in turn determines the charge $Q$, pressure $P$, and coupling constants $\alpha_i$. The parameter space admitting four phases is small, but is not a set of measure zero.
\begin{figure}
    \centering
	\includegraphics[width=0.44\textwidth, trim = 0cm -4.5cm 0cm 0cm]{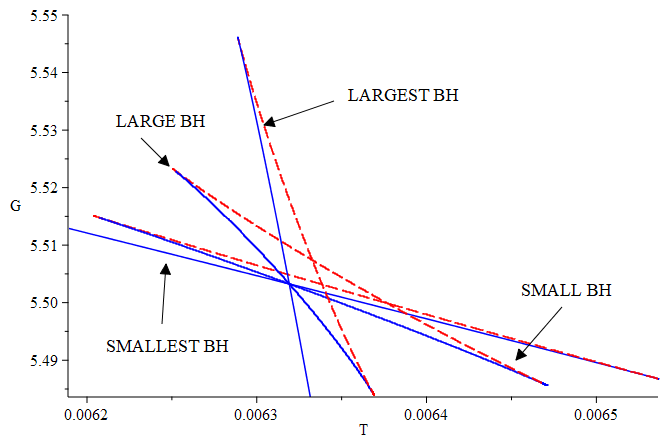}
	\includegraphics[width=0.52\textwidth]{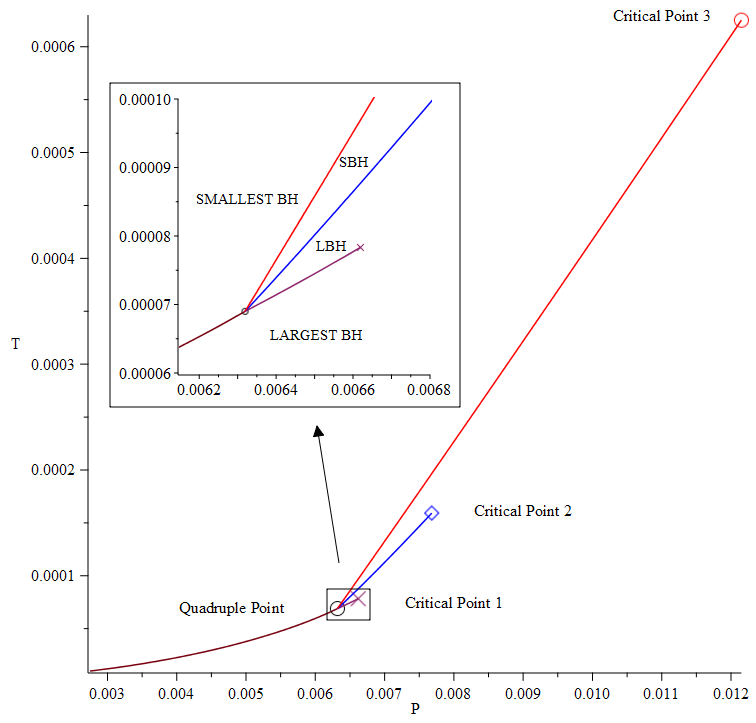}

	\caption{\textbf{Quadruple point.} $P = 6.9 \times 10^{-5}, Q = 6.75112, \alpha_2 = -20.63286635, \alpha_3 = 1379.056050, \alpha_4 = -133263.0329, \alpha_5 = 1.550137197\times 10^7, \alpha_6 = -2.017480713 \times 10^9, \alpha_7 = 2.831663674 \times 10^{11}.$ Coupling constants are rounded to 10 digits. \textit{Left.} Three swallowtails in the Gibbs free energy overlap at a single point. Dashed lines indicate negative specific heat. \textit{Right.} P-T phase diagram showing four distinct phases separated by first order phase transitions merging at the quadruple point. 
	}
	\label{fig:8horizonquad} 
\end{figure}

Further adjusting the locations of the extrema, the three inflections can be made to occur at the same temperature, resulting in the quadruple point shown in Fig.~\ref{fig:8horizonquad}. For high pressures and temperatures, only one black hole phase is observed. As the temperature and pressure are lowered, new phases emerge at distinct  critical points, separated by first order phase transitions. In the pressure range $P\in ( 6.90 \times 10^{-5}, 7    .83 \times 10^{-5})$, four distinct black hole phases exist,  characterised by their sizes. As the pressure is lowered to $6.90 \times 10^{-5}$, the four phases merge at a critical point, and only the largest and smallest phases exist at lower pressures.

\begin{figure}
    \centering
	\includegraphics[width=0.72\textwidth]{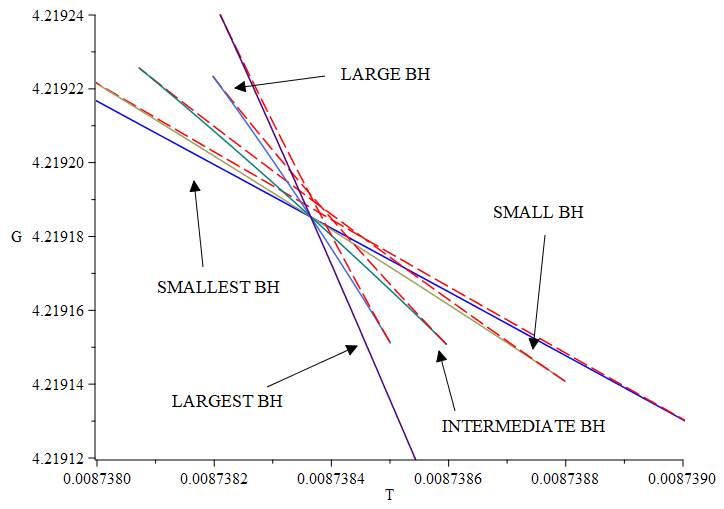}

	\caption{\textbf{Quintuple point.} G-T plot of a black hole quintuple point obtained with the first method using seven coupling constants. The extrema of $T(r_+)$ are chosen to be at $r_+=3,\ 3.1,\ 5.29,\ 5.5,\ 5.85,\ 6.39505,\ 7.15,\ 8.14996,\ 9.3,\ 10.38$. Dashed lines indicate negative specific heat. 
	}
	\label{fig:10horizonquin} 
\end{figure}

Multi-critical points of higher order can likewise be achieved by introducing more coupling constants. In general, two additional coupling constants are needed for each new $n$-tuple point; for example two coupling constants ($\alpha_2$ and $\alpha_3$) are needed to obtain  a triple point.
For an $n$-tuple point, the Gibbs free energy needs to support $n-1$ swallowtails, indicating $n-1$ first order phase transitions. Swallowtails are constructed by choosing where the extrema of $T(r_+)$ occur; namely solving
\begin{equation}
    T^\prime (r_{+}^{(1)})=...=T^\prime (r_{+}^{(2(n-1))})=0
\end{equation} 
for $P, Q, \alpha_i$, where the prime denotes the derivative with respect to $r_+$, as each pair of local minima and   maxima produce a swallowtail in the Gibbs free energy. Once a sufficient number of swallowtails are obtained, it is possible to make additional adjustments  to the locations of the extrema so that the inflections of $T(r_+)$ occur at the same temperature, forming $n-1$ swallowtails that overlap in the Gibbs free energy. Fig.~\ref{fig:10horizonquin} shows a quintuple point realized using this method.

An alternate method for finding  multi-critical points is to choose values of $r_+$ where a line of constant temperature $T^*$ intersects $T(r_+)$. An $n$-tuple point has $2n-1$ such intersects, obtained by solving 
\begin{equation}
    T(r_+^{(1)}) = ... = T(r_+^{(2n-1)}) = T^*
\end{equation}
for $T^*, P, Q, \alpha_i$. This yields $n-1$ swallowtails in the Gibbs free energy that only require slight adjustments to merge. This approach can more easily generate multi-critical points using the minimum number of coupling constants; using it we can obtain  a quadruple point with $\alpha_i \neq 0$
for $i \leq 5$, two fewer couplings  than for the quadruple point shown in 
Fig.~\ref{fig:8horizonquad}.

A truncation of the Power-Maxwell theory at finite $N$ also yields black hole solutions with multicritical behaviour.  For example we can
find a quadruple point for $N=5$ Power Maxwell theory.  Solving \eqref{Max5} for $\Phi^\prime(r)$ and then inserting the result into \eqref{Ein5}    yields the solution
\begin{align}
U(r) &= \frac{r^2}{l^2}+1-\frac{2M}{r} \nonumber \\
&\quad +  \frac{1}{r} \int^r d\tilde{r} \left(\tilde{r}^2 \sum_{n=1}^5 \left(n-\frac{1}{2}\right)\alpha_n \left( -2 (\Phi(\tilde{r})^\prime)^2\right)^n\right)
\end{align}
where $(\Phi^\prime)$ is determined by solving \eqref{Max5}. 
 Analytic solutions cannot be obtained in general, but series inversion
yields $(\Phi^\prime)$ as a power series in  $-\frac{Q}{r^2}$  with the leading term linear in this quantity and all coefficients in the series uniquely determined in terms of $\{\alpha_2,\ldots,\alpha_5\}$ (setting $\alpha_1=1$).  The metric function $U(r)$ likewise becomes an infinite series in
$\frac{Q^2}{r^2}$, with all coefficients similarly uniquely determined.  
 It is straightforward to show that $M$ and $Q$ can be respectively interpreted as the mass and charge of the black hole. 
 
\begin{figure}
    \centering
	\includegraphics[width=0.65\textwidth]{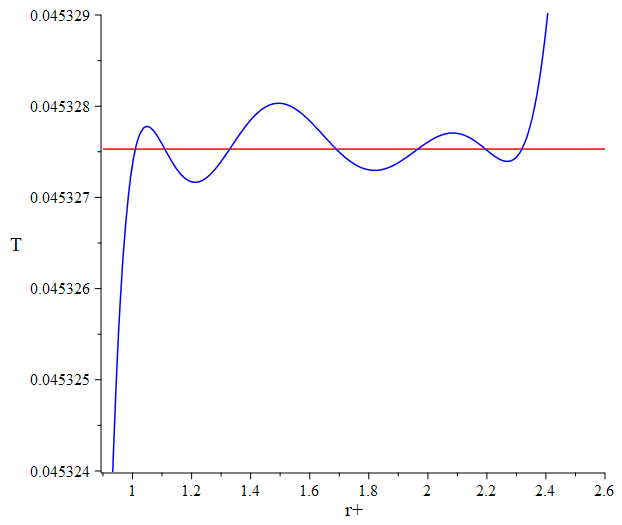}	
	\caption{\textbf{Quadruple point for $N=5$ Power Maxwell Theory}  The single value $T^*$ (here $T^* =0.0453275294319341$) intersecting
all inflection points of the $T(r_+)$ curve indicates the presence of a quadruple point. In this example
$\Lambda = -0.08987715997, Q = 0.8891866017, \alpha_2 = 2.276862279, \alpha_3 = 15.37997040, \alpha_4 = 35.22356894, \alpha_5 = 47.91591656$.
	}
	\label{fig:sm} 
\end{figure}

 It is not necessary to explicitly compute this series to understand the thermodynamic behaviour and demonstrate the existence of a quadruple point.
Since $T = \frac{U^\prime(r_+)}{4\pi}$, we obtain 
\begin{equation}
T  = \frac{r_+^2 \sum_{n=1}^5 \left(n-\frac{1}{2}\right)\alpha_n \left( -2 (z_+)^2\right)^n - r_+^2\Lambda +1}{4\pi r_+}
\end{equation}
from  \eqref{Ein5}, where
\begin{equation}
 \frac{1}{2} r_+^2 \sum_{n=1}^5  n \alpha_n \left( -2 z_+^2\right)^n - Q z_+ = 0
 \end{equation}
 from \eqref{Max5} , setting $z_+ = \Phi^\prime(r_+)$.  It is straightforward to choose a set of values for the parameters $\{P,Q,\alpha_2,\ldots,\alpha_5\}$ that yield a set of three swallowtails, corresponding to four distinct black hole phases.  Small adjustments of these parameters then yield a solution $T^*$ at which all inflection points 
in the $T(r_+)$ curve  occur at the same temperature, signifying a quadruple point.   The values of these parameters can be obtained numerically.
An example is given in  Fig.~\ref{fig:sm}, where $T^* =0.0453275294319341$.  Note that since
\be
dG=VdP + \Phi dQ + \sum_{i=2}^N \mathcal{B}_i d\alpha_i - SdT,
\ee
the critical points of $G(r_+)$ coincide with those of $T(r_+)$, and thus the behaviour of $G(T)$ can be understood by analyzing $T(r_+)$ alone.
 
 Any given odd value of $N=2n-1$ in Power-Maxwell theory will yield at most an $(n+1)$-critical point, in agreement with the required number of thermodynamic conjugate pairs highlighted in the main text. These maximally multicritical points, as well as those of lower dimensionality,   can be obtained using methods similar to those given above.

By generalizing the scaling  arguments used to obtain the Smarr relation \cite{Kastor:2009wy}, we can regard each $\alpha_i$ as a thermodynamic variable. Noting
that dimensionally $[M]=L, [S]=L^2, [P]=L^{-2}, [Q]=L, [\alpha_i]=L^{2(i-1)}$, it is straightforward to show
that the first law
\begin{equation}
    dM =T dS +V dP +\Phi dQ + \sum_{i=2}^n \mathcal{B}_i d\alpha_i.
\end{equation}
and   generalized Smarr relation
\begin{equation}
    M=2\left(TS - PV+ \sum_{i=2}^n (i-1) \mathcal{B}_i \alpha_i \right) + \Phi Q.
\end{equation}
both hold for the solutions we obtain, where
\begin{equation}
    \mathcal{B}_i = \frac{\partial M}{\partial \alpha_i}
\end{equation}
is thermodynamically conjugate to the $\alpha_i$ coupling.  These quantities  play a role similar to polarization susceptiblities in non-linear   optics, generalizing the notion of vacuum polarization in Born-Infeld Electrodynamics  
\cite{Gunasekaran:2012dq}.  We note that Born-Infeld Electrodynamics has only a single additional thermodynamic parameter compared to Einstein-Maxwell theory, which is an insufficient number  for multicritical behaviour to be realized.

\section{Gibbs Phase Rule for Black Holes}
 
Our results indicate that black holes can behave like multicomponent chemical systems seen elsewhere in nature \cite{PhysRevLett.125.127803,Akahane2016,Garcia2017}. In such systems
the generalized Gibbs phase rule \cite{Sun:2021gpr} relates the number of coexistent phases $\textsf{P}$ to the number of thermodynamic conjugate pairs
$\textsf{W}$ via
\begin{equation}\label{GPR}
    \textsf{F}=\textsf{W}-\textsf{P}+1
\end{equation}
where  
setting $\textsf{W}=\textsf{C}+1$  recovers the Gibbs phase rule for simple systems, with $\textsf{C}$ the number of constituents in a multicomponent chemical system. Modern functional materials are not simple,  having additional degrees of freedom that can do thermodynamic work represented by the more general quantity $\textsf{W}$ \cite{Sun:2021gpr}.  
The number of degrees of freedom
$\textsf{F}$ is the number of independent intensive parameters.   
$n$-tuple points are marked by simplices connecting $n$ phases on the internal energy surface in terms of the extensive thermodynamic variables $U(S,V,...)$, and $\textsf{F}$ is determined by the nullity of the set of points in $\mathbb{R}^n$ defining the simplex \cite{Sun:2021gpr}. The maximum number of coexistent phases is $\textsf{P}=\textsf{W}+1$, attained for $\textsf{F}=0$, as a set of $\textsf{W}+1$ affinely independent points in $\mathbb{R}^{\textsf{W}+1}$ has 0 nullity. 

The situation is somewhat different in Black Hole Chemistry. A charged AdS black hole has only two phases ($\textsf{P}=2$), large and small, respectively analogous to the high-entropy liquid and low-entropy gas phases of a Van der Waals fluid \cite{Kubiznak:2012wp}. The analogue of the lowest-entropy solid phase is thermal AdS \cite{Kubiznak:2014zwa}, which cannot be attained due to charge conservation. If charge vanishes, this phase exhibits a first order phase transition with a higher-entropy large black hole \cite{Hawking:1982dh}
analogous to a solid-liquid transition  \cite{Kubiznak:2014zwa}
but is missing the analogous  highest entropy gaseous phase, again giving $\textsf{P}=2$.  Consequently neither the Schwarzschild-AdS nor Reissner-Nordstrom-AdS black holes can exhibit triple points and $\textsf{F}=2$. 

More generally $2n-1$  thermodynamic conjugate pairs 
allow for $n$-tuple phase transitions. Triple points have been observed in black hole systems in $D=4$ 
but with at least five thermodynamic conjugate pairs \cite{Zhang:2020obn}. In higher dimensions, triple points can exist with fewer conjugate pairs \cite{Altamirano:2013uqa,Wei:2014hba,Frassino:2014pha}, but still within bounds of the generalized Gibbs' phase rule. By regarding the coupling constants in Power-Maxwell theory as extensive thermodynamic variables, we find that each new phase requires two additional thermodynamic conjugate pairs.  The generalized Gibbs phase rule indicates that an $n$-tuple point in non-linear electrodynamics has $\textsf{F}=n$ degrees of freedom. 
In general other black hole systems with a sufficiently large number of parameters can be expected to exhibit multicritical behaviour provided the dependence of the temperature on the horizon radius is sufficiently complicated. 
The implications of this for the microstructure of black holes \cite{Wei:2019yvs,Wei:2019uqg,Wei:2021krr} remains to be understood. 

Schwarzschild-AdS black holes, however, can exhibit 
radiation-small-large 
triple points  
with only three conjugate pairs in higher curvature gravity
\cite{Hull:2021bry}, fully analogous to the triple point of water, provided their horizon geometry does not have constant curvature.  This suggests that the Gibbs Phase rule \eqref{GPR} can be satisfied with $\textsf{F}=0$ for an $n$-tuple critical point for at least some gravitational systems. Elucidation of the necessary and sufficient conditions to attain a  black hole multicritical point remains an interesting open question.

\section*{Acknowledgements}
\label{sc:acknowledgements}

This work  supported in part by the Natural Sciences and Engineering Research Council of Canada (NSERC).  Perimeter Institute and the University of Waterloo are situated on the Haldimand Tract, land that was promised to the Haudenosaunee of the Six Nations of the Grand River, and is within the territory of the Neutral, Anishnawbe, and Haudenosaunee peoples. We are grateful to Bill Power, David Yevick, and Donna Strickland for helpful discussions.

\bibliographystyle{JHEP}

\providecommand{\href}[2]{#2}\begingroup\raggedright\endgroup

\end{document}